\documentclass[prl,reprint,amsmath,amssymb,aps,superscriptaddress,longbibliography]{revtex4-2}
\usepackage{graphicx}
\usepackage{dcolumn}
\newcolumntype{d}[1]{D{.}{.}{#1}}
\usepackage{bm}
\usepackage[colorlinks,bookmarks=false,citecolor=blue,linkcolor=red,urlcolor=blue]{hyperref}
\usepackage{times}
\usepackage{braket}
\graphicspath{{Figs/}}
\usepackage{soul}

\newcommand{\SU}{\ensuremath{\mathrm{SU}}}

\begin{document}
\title{Extracting the Speed of Light from Matrix Product States}
\author{Alexander A. Eberharter}
\affiliation{Institut f\"ur Theoretische Physik, Universit\"at Innsbruck, A-6020 Innsbruck, Austria}
\author{Laurens Vanderstraeten}
\affiliation{Department of Physics and Astronomy, University of Ghent, Belgium}
\affiliation{Center for Nonlinear Phenomena and Complex Systems, Université Libre de Bruxelles, Brussels, Belgium}
\author{Frank Verstraete}
\affiliation{Department of Physics and Astronomy, University of Ghent, Belgium}
\affiliation{Department of Applied Mathematics and Theoretical Physics, University of Cambridge, Cambridge, United Kingdom}
\author{Andreas M. Läuchli}
\affiliation{Laboratory for Theoretical and Computational Physics, Paul Scherrer Institute, 5232 Villigen, Switzerland}
\affiliation{Institute of Physics, \'{E}cole Polytechnique F\'{e}d\'{e}rale de Lausanne (EPFL), 1015 Lausanne, Switzerland}
\date{\today}

\begin{abstract}
We provide evidence that the spectrum of the local effective Hamiltonian and the transfer operator in infinite-system matrix product state simulations are identical up to a global rescaling factor, i.e.~the speed of light of the system, when the underlying system is described by a 1+1 dimensional CFT. We provide arguments for this correspondence based on a path integral point of view. This observation turns out to yield very precise estimates for the speed of light in practice, confirming exact results to high precision where available, but also allowing us to finally determine the speed of light of the non-integrable, critical $SU(2)$ Heisenberg chains with half-integer spin $S>1/2$ with unprecedented accuracy. We also show that the same technology applied to doped Hubbard ladders provides highly accurate velocities for a range of dopings. Combined with measurements of compressibilities we present new results for the Luttinger liquid parameter in the Luther-Emery regime of doped Hubbard ladders, outperforming earlier approaches based on the fitting of real-space correlation functions. 

\end{abstract}
\maketitle

\paragraph{Introduction ---}
Relativistic invariance is a property that quantum many body phases or critical points often exhibit in the low-energy or long-wavelength limit. The speed of light (or speed of sound) translates between momentum and energies $E(k)=v|k|$ for massless systems ($\hbar\equiv 1$). This family of systems encompasses Goldstone phases corresponding to the symmetry breaking of continuous symmetries, and quantum critical points in low dimensions with a dynamical critical exponent $z=1$, in particular conformal field theories (CFTs). 

The speed of light enters many important quantities in quantum many body systems, such as the Casimir effect of the ground state energy density~\cite{Bloete1986,Affleck1986}, the low-temperature behaviour of the specific heat~\cite{Bloete1986,Affleck1986}, formulas for the critical exponents of Luttinger liquids~\cite{Giamarchi2003}, or the prefactor in the finite-size energy spectrum of conformal field theories, to name a few. In the past an accurate numerical extraction of the velocity of a non-integrable quantum many body system was rather tedious and often not very precise. For example, before the discovery of the relation between the entanglement entropy and the central charge, the latter was estimated using an analysis of the Casimir correction of the ground state energy density. In that formula  the speed of light also enters, and estimates of the velocity were notoriously difficult, thus hampering also the precision of the extracted central charge. In recent years, using quantum Monte-Carlo \cite{Sen2015,Tan2017} or tensor-network methods \cite{Haegeman2012, Vanderstraeten2019b}, dispersion relations can be computed with some precision, but reaching high accuracy on the velocity for strongly correlated models remains challenging.

On a different line of research, the study of 1+1 dimensional CFTs using entanglement measures has become a very active field. Over time it has been demonstrated or observed that in many places glimpses of the CFT appear. The entanglement entropy of a subsystem is governed by the central charge $c$ \cite{Calabrese2004} and the entanglement spectrum is described by the spectrum of an appropriate boundary CFT~\cite{Laeuchli2013,Cardy2016}. This spectrum can also be observed in the corner transfer matrix of two-dimensional classical statistical mechanics models \cite{Baxter1982}.

The critical ground-state properties of one-dimensional lattice models can be simulated accurately using the formalism of (infinite) matrix product states (MPS). Although the finite bond dimension of the MPS always induces a cutoff on the entanglement in a critical model, the critical data can be extracted from a finite-entanglement scaling that is ruled by the properties of the underlying CFT. For example, the scaling of the entanglement entropy was shown to be solely determined by the central charge \cite{Nishino1996b, Tagliacozzo2008, Pollmann2009, Pirvu2012}. In this finite-entanglement scaling, the MPS transfer matrix seems to play an important role, as it encodes the structure of all effective length scales \cite{Rams2018, Vanhecke2019} and correlation functions, including the structure of the low-lying excitations in the system \cite{Zauner2015}. On the other hand, for finite-size systems the local effective Hamiltonian was shown to be identical to the spectrum of the finite-size spectrum \cite{Chepiga2017}, which yields a very powerful method for extracting critical data \cite{Chepiga2020}. Nonetheless, the detailed structure of both the transfer matrix and the effective Hamiltonian in the regime of finite-entanglement scaling are not understood.

In this paper we now provide strong evidence that these two spectra are in fact identical up to a global rescaling factor, which we identify as the speed of light of the system when the underlying system is described by a 1+1 dimensional CFT. Apart from its importance as a conceptual result, this finding offers a long-sought and very precise approach to measure the speed of light in MPS simulations.

\begin{figure}
    \centering
    \includegraphics{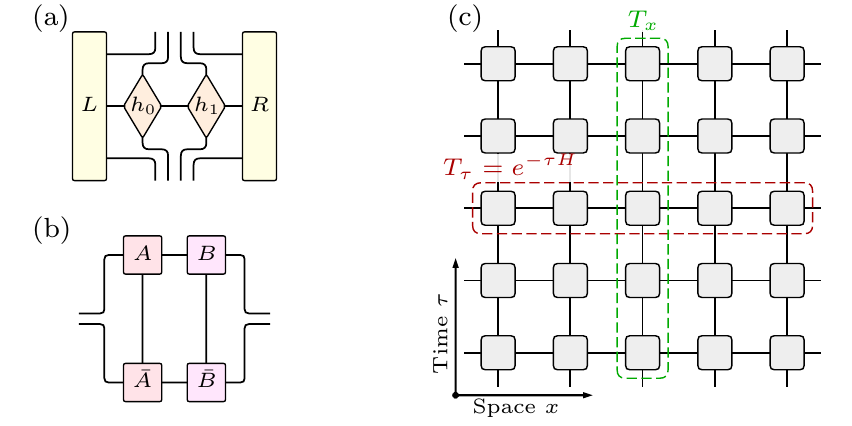}
    \caption{Tensor network diagrams for the effective Hamiltonian (a) and the MPS transfer matrix (b) for an infinite MPS consisting of a two-site unit cell formed by the tensors $A$ and $B$. $h_0$ and $h_1$ are the tensors of the MPO which represents the Hamiltonian, whereas $L$ and $R$ are the left and right boundary tensors of the MPS. (c) The MPS path-integral representation with temporal and spatial transfer matrices (for a one-site unit cell); the effective Hamiltonian and MPS transfer matrix appear as truncated versions of these two transfer matrices.}
    \label{fig:diagram}
\end{figure}

\paragraph{Infinite system matrix product states ---}

In the following we study one-dimensional translation invariant lattice systems with Hamiltonian $H$, which we assume can be presented by a matrix product operator (MPO) \cite{Pirvu2010}. For illustration, we consider the situation that the ground state of $H$ can be approximated by an infinite MPS with a two-site unit cell ($l_\mathrm{UC}=2)$, with $A$ and $B$ the two MPS matrices. Following standard MPS notation and algorithms \cite{McCulloch2008, Kjaell2013, Hauschild2018, ZaunerStauber2018, Vanderstraeten2019} we study the spectrum of the effective Hamiltonian $\tilde{H}$, as shown in Fig.~\ref{fig:diagram}(a). The effective Hamiltonian $\tilde{H}$ is a Hermitian matrix of linear dimension $D^2 d^2$, with $D$ the MPS bond dimension and $d$ the local dimension of the physical sites~\footnote{Note that we could also consider the effective Hamiltonian for the single site or zero-site updates \cite{Vanderstraeten2019}, for which the dimension would be $D^2 d$ or $D^2$.}. While in a standard DMRG-like update step only the ground state energy and corresponding eigenvector of $\tilde{H}$ are needed for a subsequent update of the MPS tensors, here we calculate a certain number of excited energies as well. We call these energies $E_i$, with $E_0$ being the ground state energy. In the presence of symmetries we can calculate excited energies in all relevant symmetry sectors.

\par In the MPS framework the transfer matrix $\tilde{T}$ of the MPS across its unit cell (see Fig.~\ref{fig:diagram}(b)) is a very powerful object, as it allows to measure the correlation length(s) and also gives access to the real space structure of correlations~\cite{Zauner2015}. The transfer operator $\tilde{T}$ is a square matrix of linear dimension~$D^2$. The largest eigenvalue of $\tilde{T}$ for a normalized infinite MPS is one, whereas all other eigenvalues of $\tilde{T}$ have absolute values $|\epsilon_i| < 1,\ i>0$. The correlations lengths are then obtained as $\xi_i= -  l_\mathrm{UC} / \ln |\epsilon_i|$, where we express the correlation lengths in lattice spacings of $H$. 

\par Our central observation is that when the Hamiltonian $H$ is tuned to a parameter set where it is governed by a 1+1 dimensional CFT at low energies, then 
\begin{equation}
    (E_i-E_0) = v / \xi_i,\quad  i>0,
\end{equation}
with $v$ the effective speed of light. We show examples of this observation below and its practical power to determine the speed of light, but want to highlight here that this property is emergent, and so far only expected to be valid in a situation described by a CFT with a single velocity. We expect this correspondence to be valid in the regime of finite-entanglement scaling, where the spacings in the spectrum of both operators are due to the finite bond dimension in the MPS approximation; for a critical model, both spectra become continuous in the limit of infinite bond dimension. We also want to stress that this novel way to determine the speed of light only requires to calculate an additional set of excited state energies of the effective Hamiltonian $\tilde{H}$ in an existing infinite MPS implementation. We suggest to first calculate a large number of eigenvalues of both $\tilde{H}$ and $\tilde{T}$ to check for the spectral correspondence in a new situation, but then in principle the subleading energy and transfer matrix eigenvalue is enough to determine the speed of light $v$ to high accuracy.

\paragraph{Path-integral Viewpoint ---}

At first sight, it seems surprising that these two operators should have the same low-energy spectrum. The first clue is the observation that the MPS transfer matrix contains crucial information about the low-energy spectrum of the Hamiltonian for which the MPS was optimized. In Ref.~\onlinecite{Zauner2015} this curious feature was understood through the path-integral representation: The ground state $\ket{\psi_0}$ of a given Hamiltonian $H$ is obtained by applying an infinite stack of MPO-representations of $e^{-H \tau}$ (with $\tau$ a small time step) onto a random initial state, so that $\braket{\psi_0|\psi_0}$ can be represented as the contraction of an infinite two-dimensional tensor network \cite{Tirrito2018}, see Fig. \ref{fig:diagram}(c). In this tensor network, we can identify the spatial transfer matrix $T_x$ as an infinite column and the temporal transfer matrix $T_\tau$ as an infinite row. When the low-energy sector of the Hamiltonian is described by a relativistic field theory, we expect that the low-energy spectra of both transfer matrices are equivalent, up to global factor that is the effective velocity in the system \cite{Zauner2015, Tirrito2018}. When approximating the ground state of $H$ as an MPS, we are effectively truncating both transfer matrices  \cite{Rams2015, Bal2016}.
\par We can understand the effect of this truncation by first considering a fully isotropic two-dimensional tensor network, corresponding to the partition function of an isotropic classical stat-mech model at criticality. The transfer matrices in the horizontal and vertical directions are now the same (with effective velocity equal to one). In the case that the MPO tensors are invariant under reflection and conjugation, the transfer matrix is hermitian and we can find an MPS of the row-to-row transfer matrix through a variational principle \cite{Fishman2018, Vanderstraeten2022}. In this case, the variational optimality condition for the MPS tensors is equivalent to a set of fixed-point equations \cite{Haegeman2017} that are manifestly rotation invariant, which directly implies that the truncated transfer matrices in the horizontal and the vertical directions are the same operators, possibly up to a gauge transformation \cite{Haegeman2017} that leaves the spectrum invariant. The fact that the truncation is isotropic in this case is also apparent in the isotropic version of the corner-transfer matrix renormalization group \cite{Baxter1978, Nishino1996}.
\par For anisotropic critical partition functions the exact transfer matrices are still expected to yield the same low-energy spectrum, now with a non-trivial effective velocity. Now the MPS fixed-point equations are no longer rotation invariant, but the truncation is still performed in an isotropic fashion, as it corresponds to the truncation of the singular values that reside in a corner tensor \cite{Haegeman2017}. This suggests that the effect of the truncation is still isotropic, at least in the low-energy sector, and that we can expect the transfer matrices in both directions to have the same low-energy spectra up to an effective velocity. Finally, we can expect that this observation continues to hold in the (anisotropic) Hamiltonian limit, where time becomes continuous \cite{Tirrito2018} and the logarithm of the spectrum of the truncated temporal transfer matrix becomes the spectrum of the effective Hamiltonian as defined above.

\begin{figure}
    \includegraphics{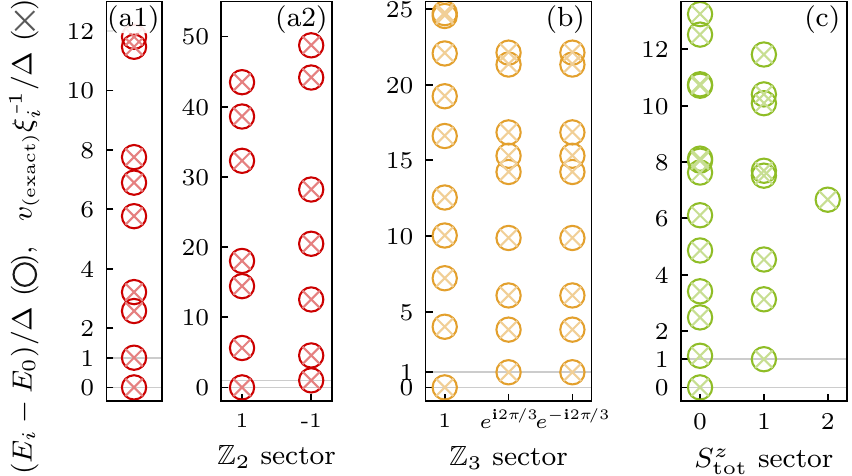}
    \caption{\label{fig:spectrum} Spectra of the effective Hamiltonian and transfer matrix of the infinite ground state MPS with bond dimension $D = 128, 256$ and $512$ for (a), (b) and (c) respectively. Both spectra differ only by a single global factor $\nu_\textrm{(exact)}$ which we take from exactly known velocities. The common vertical axis is scaled using the effective Hamiltonian's energy gap $\Delta$, such that the first non-zero eigenvalue is set to one.}
\end{figure}

\paragraph{Applications to Spin Chains ---}

In Fig.~\ref{fig:spectrum} we present our central observation, the compelling correspondence between the low energy spectrum of the effective Hamiltonian $\tilde{H}$ as circles and the inverse correlation lengths derived from the transfer matrix eigenvalues as crosses, where we use the exactly known velocities $v_\mathrm{(exact)}$ for a parameter-free match of the spectra. In panel (a1) and (a2) we show data for the critical transverse-field Ising model, without (a1) and with (a2) imposing $\mathbb{Z}_2$ symmetry in the MPS tensors. In panel (b) we show $\mathbb{Z}_3$ symmetry resolved spectra for the critical three state quantum Potts chain. Finally in (c) we display the $S^z_\mathrm{tot}$ resolved spectra for the $S=1/2$ antiferromagnetic Heisenberg chain. All the spectra show an almost perfect correspondence between energy and transfer matrix spectrum. The range of matching eigenvalues depends on the bond dimension $D$ of the MPS: For larger bond dimension a larger range of eigenvalues match before deviations become visible.

We can now use the first subleading eigenvalue in each spectrum type to directly determine the velocity from the numerical data $v_\mathrm{MPS}=(E_i-E_0) \times \xi_i$ and display the results in Tab.~\ref{tab:velocity} \footnote{The symbol $J$ is used for the nearest neighbor coupling strength of the Potts and Heisenberg models}. The determined velocity for the critical Ising model is about three orders of magnitude more accurate than previous numerical approaches~\cite{Chepiga2017,Wu2020}, and about two orders of magnitude more accurate for the $Q=3$ and $Q=4$ Potts models~\cite{Wu2020}.

\begin{table}
    \begin{ruledtabular}
    \begin{tabular}{l|d{4} l | d{4} l | d{3} l}
             \multicolumn{1}{c|}{$Q$} & \multicolumn{2}{c|}{2} & \multicolumn{2}{c|}{3} & \multicolumn{2}{c}{4}\\
            \hline
            $v$ (exact) & 2& & 2.598076 & $\ \approx \scriptstyle \, 3\sqrt{3}/2$ & 3.141593& $\ \ \ \approx \pi$\\
            $v \ (D)$ & 1.999975&(128) & 2.59799&(512) & 3.14141&(1024) \\
            $v \ (D)$ & 1.999904&(64) & 2.59785&(256) & 3.14116&(512)
        \end{tabular}
    \end{ruledtabular}
    \medskip
    \begin{ruledtabular}
        \begin{tabular}{l|d{5}l | d{4} | d{4} | d{3}}
             \multicolumn{1}{c|}{Spin $S$} & \multicolumn{2}{c|}{1/2}
             & \multicolumn{1}{c|}{3/2} & \multicolumn{1}{c|}{5/2} & \multicolumn{1}{c}{7/2}\\
            \hline
            $v$ (exact) & 1.570796& $\ \approx \pi/2$ & \multicolumn{1}{c|}{-} & \multicolumn{1}{c|}{-} & \multicolumn{1}{c}{-} \\
            $v \ (D=1024)$ & 1.57059&($D=512$) & 3.4392 & 5.4037 &7.386\\
            $v \ (D=512)$ & 1.57035&($D=256$) & 3.4379 & 5.4014 &7.377\\
        \end{tabular}
    \end{ruledtabular}
    \caption{\label{tab:velocity} 
    Comparison of numerical results of the velocities for the two highest bond dimensions $D$ used and the exact values if available
    ~\cite{Alcaraz1988,DesCloizeaux1962}. (Top) $Q$-state Potts model, (bottom) SU(2) AF model. Here $J=1$ is used.
    }
\end{table}

\begin{figure}
    \includegraphics{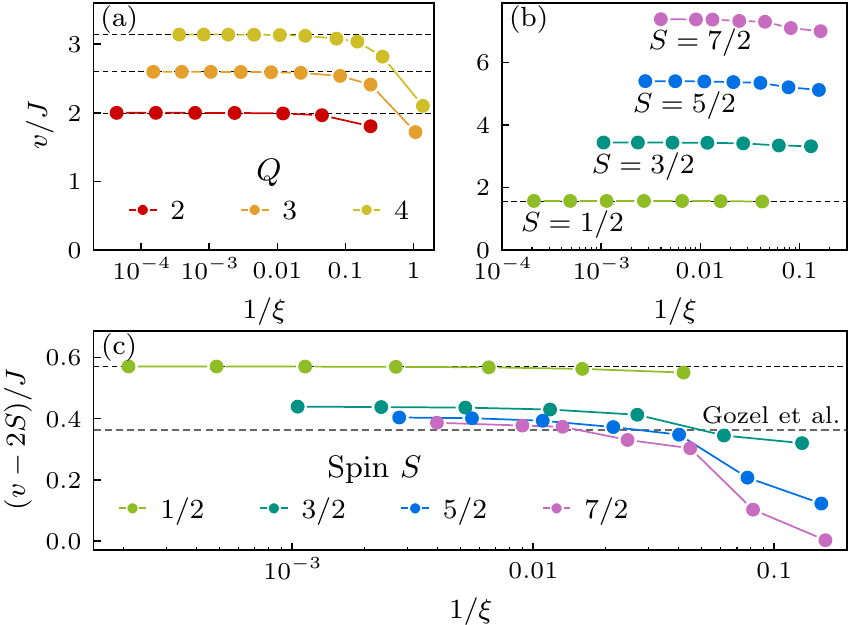}
    \caption{\label{fig:v_potts_af} The velocity obtained by our method as $\nu = -(E_1 - E_0)/\log(|\epsilon_1|)$ plotted against the inverse of the correlation length $\xi^{-1} = -\log(|\epsilon_1|)$ for several values of bond dimension $D$. (a) Critical Potts models for $Q=2$ (Ising)$,3,4$. (b) Critical $SU(2)$ AF Heisenberg chains. Light dashed lines denote exact results from integrability. The bolder dashed line in (c) is taken from Ref.~\cite{Gozel2019}.}
\end{figure}

As a nontrivial application we now want to check a long-standing analytical prediction~\cite{Haldane1983} and its recent refinement~\cite{Gozel2019} for the speed of light (spin wave velocity) of the half-integer spin-$S$ Heisenberg chains, which flow to the $SU(2)_1$ Wess-Zumino-Novikov-Witten CFT in the infrared with central charge $c=1$ \cite{Affleck1987}. The speed of light is not universal and is not accurately known beyond the $S=1/2$ case \cite{Hallberg1995}, but from a mapping onto the non-linear $\sigma$ model one expects the velocity to scale as $v(S)\approx J(2S + |(2/\pi-1)|)$ for large $S$~\cite{Gozel2019}. 
In Fig.~\ref{fig:v_potts_af}(b,c) we make use of our new method to extract the velocities for several $S$. In panel (b) we display the velocities on a common scale and one can see how the velocities grow linearly with $S$. We plot the velocity as a function of the correlation lengths in the MPS, which are induced by the different bond dimensions. For large bond dimensions the velocities converge to a constant value, see also Tab.~\ref{tab:velocity}. In Fig.~\ref{fig:v_potts_af}(c) we plot $(v_\mathrm{MPS}(S)-2SJ)/J$, so that we can check that indeed the velocities for larger $S$ and correspondingly large bond dimension approach the positive offset $|(2/\pi-1)|\approx 0.36338$.

\paragraph{Application to Luther-Emery Liquids ---}

Next we test our method for systems that fall within the Luther-Emery universality class, characterized by gapped spin excitations and a single gapless charge mode. Obtaining the infrared properties of Luther-Emery liquids has become very relevant in the context of MPS-based simulations of two-dimensional $t{-}J$ and Hubbard models \cite{Jiang2018, Jiang2020, Jiang2020b, Jiang2021, Gong2021, He2021}. Getting accurate estimates of critical exponents is crucial in these studies, but requires the fitting of the correlation functions on long cylinders, the use of large bond dimensions and tedious extrapolation procedures. Here we show that using infinite MPS and our velocity scaling yields a more efficient method for obtaining accurate exponents.
\par We will consider Hubbard models \cite{Essler2005} with nearest-neighbour hopping (with parameter $t$) and an on-site interaction strength $U$. We also introduce a chemical potential $\mu$ in order to consider incommensurate electron fillings between $n=0$ (zero filling) and $n=2$ (complete filling). Consequently we cannot exploit charge conservation, so all simulations are performed with $\SU(2)$ spin-rotation symmetry and fermion parity in the MPS representation.
\par
\begin{figure}
\center{\includegraphics{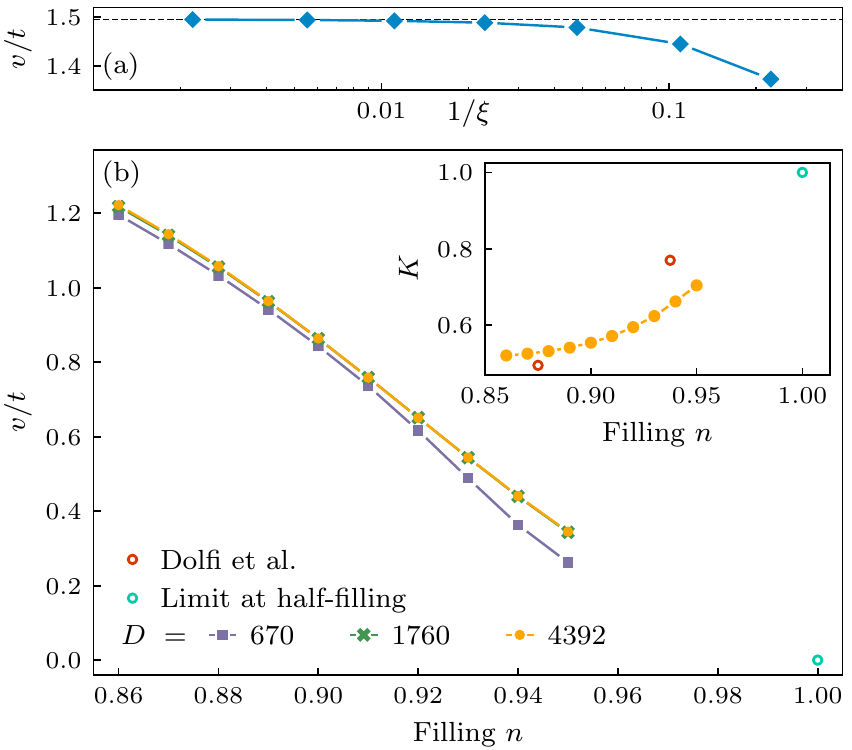}}
\caption{Numerical results for the velocity of (a) the Hubbard chain with attractive interaction $U/t = -2$ and (b) the Hubbard ladder with repulsive interaction $U/t=8$. For the chain we set $\mu/t = -1.5$ resulting in a filling of $n\approx0.78$ and repeat the MPS ground state calculations for multiple values of the bond dimension in the range $D=23$ to $1321$. Here we show the effective velocity plotted against inverse correlation length and compare our data with the exact Bethe-ansatz result. For the ladder we plot the velocity against the filling $n$ for three values of the bond dimension. In the inset we show the Luttinger parameter extracted from the velocity and the compressibility for the largest bond dimension, and compare to the result in Ref.~\onlinecite{Dolfi2015} and the limiting value at half filling.}
\label{fig:hubbard}
\end{figure}

 We start with the simple Hubbard chain with an attractive interaction, $U<0$. The ground state exhibits a spin gap and critical charge correlations~\cite{Essler2005}, and falls within the Luther-Emery class. We perform ground state optimizations at different bond dimensions, and extract the effective velocity through the transfer matrix and effective Hamiltonian gaps. In Fig.~\ref{fig:hubbard} we plot the velocity as a function of the inverse correlation length. We observe a quick convergence to the exact result from the Bethe ansatz \cite{Essler2005}.
\par As a more challenging case we take the (non integrable) Hubbard ladder. We work in the case of isotropic hopping and repulsive interactions $U/t=8$ and tune the chemical potential in the weakly-doped regime. In this regime, analytical results in the weak coupling limit \cite{Balents1996} suggest that there is one gapless charge mode and a gap in the spin sector, which was confirmed by early DMRG calculations \cite{Noack1994, Noack1996}. The infrared properties are determined by the Luttinger parameter $K$ \footnote{We use the convention for $K$ from Ref. \cite{White2002}, which differs by a factor of two from the one in, e.g., Ref. \cite{Dolfi2015}.}. It is known \cite{Balents1996, Schulz1999, Siller2001} that in the limit of half filling ($n\to1$) we find $K\to1$, but the numerical determination for the doped case is very challenging \cite{Dolfi2015}. Using Luttinger liquid theory \cite{Giamarchi2003}, we can circumvent the fitting of correlation functions by obtaining $K$ as the product of the velocity $v$ and the compressibility (with an additional constant that depends on the convention) \cite{White2002}, where the latter can be simply determined from the filling as a function of chemical potential.
\par Our numerical results are presented in Fig.~\ref{fig:hubbard}. We find that the extracted velocities are converged already at reasonably small bond dimensions. In the inset, we show the Luttinger parameter and compare to estimates from fitting correlation functions using finite size DMRG simulations \cite{Dolfi2015}. The latter yields estimates that depend quite strongly on the procedure for extrapolating the correlation functions to infinite MPS bond dimension, so the agreement seems reasonable.

\paragraph{Outlook ---}

In this work we have introduced the intriguing equivalence between the low-energy spectra of the transfer matrix and effective Hamiltonian in the infinite MPS simulation of critical 1+1 dimensional quantum chains, and we have illustrated the power of this equivalence for estimating the effective velocity in these systems. Our observation opens the door to the investigations of other systems of current interest, such as the nature of phase transitions in Rydberg blockaded spin chains~\cite{Rader2019,Chepiga2019}, where extended floating phases with $c=1$ and finite velocity can terminate in phase transitions with $z=2$ as the velocity $v\rightarrow0$. A more challenging scenario to explore are effective Hamiltonian and transfer matrix spectra in systems with two or more different velocities, for example a doped Hubbard chain with $c=2$, but different charge and spin velocities~\cite{Essler2005}. Finally, similarly powerful tensor network methods for 2+1 dimensional CFTs would be welcome, as Quantum Monte Carlo methods - which work well for some classes of systems~\cite{Sen2015,Tan2017} - can face difficulties in situations where the finite size spectrum is affected by interaction effects~\cite{Tang2018, Hesselmann2019, Schuler2021}. Conceptually, this work will allow for a better understanding of the structure of the MPS transfer matrix spectrum, which remains a fundamental open question within the program of finite-entanglement scaling with tensor networks.

\paragraph{Acknowledgements ---}
The authors would like to thank N.~Chepiga, T.~Giamarchi, S.~Gozel, J.~Haegeman, R.~Huang, L.~Tagliacozzo and B.~Vanhecke for inspiring discussions and J.~De Nardis and F.~Essler for sharing useful Bethe-ansatz data. AEE and AML acknowledge support by the Austrian Science Fund (FWF) through project I-4548. LV is supported by the FRS-FNRS (Belgium). This work was also supported by a grant from FWO, no. G0E1820N. Some of the simulations have been performed using the TenPy package~\cite{Hauschild2018}.

\bibliography{velocity}

\end{document}